%% file: main_arXiv.tex
\documentclass[aps,twocolumn,showpacs,amsmath,preprintnumbers,amssymb,superscriptaddress,reprint]{revtex4-2}

\usepackage{epsf}
\usepackage{graphicx}
\usepackage{color}
\usepackage[normalem]{ulem}
\usepackage{newfloat}
\usepackage[latin1]{inputenc}
\usepackage{comment}
\usepackage{amsmath}
\usepackage{physics}
\usepackage{float}

\usepackage{hyperref}

\begin{document}
\title{Tunable competing optical excitation pathways\\ in the topological surface states of Bi$_2$Te$_3$}

\author{Shin Yokoyama}
\affiliation{Graduate School of Advanced Science and Engineering, Hiroshima University, Higashi-Hiroshima, Hiroshima 739-8526, Japan}
\affiliation{International Institute for Sustainability with Knotted Chiral Meta Matter (WPI-SKCM$^2$), Hiroshima University, Higashi-Hiroshima, Hiroshima 739-8526, Japan}

\author{Takahito Takeda} 
\affiliation{Graduate School of Advanced Science and Engineering, Hiroshima University, Higashi-Hiroshima, Hiroshima 739-8526, Japan}
\affiliation{Department of Chemical System Engineering, The University of Tokyo, Bunkyo-ku, Tokyo 113-8656, Japan}

\author{Nagi Suzuki} 
\affiliation{Graduate School of Advanced Science and Engineering, Hiroshima University, Higashi-Hiroshima, Hiroshima 739-8526, Japan}
\affiliation{International Institute for Sustainability with Knotted Chiral Meta Matter (WPI-SKCM$^2$), Hiroshima University, Higashi-Hiroshima, Hiroshima 739-8526, Japan}

\author{Takuma Iwata} 
\affiliation{Graduate School of Advanced Science and Engineering, Hiroshima University, Higashi-Hiroshima, Hiroshima 739-8526, Japan}
\affiliation{International Institute for Sustainability with Knotted Chiral Meta Matter (WPI-SKCM$^2$), Hiroshima University, Higashi-Hiroshima, Hiroshima 739-8526, Japan}

\author{Ping Zhou} 
\affiliation{Faculty of Physics and Center for Nanointegration (CENIDE), University of Duisburg-Essen, 47057 Duisburg, Germany}

\author{Yogendra Kumar} 
\affiliation{Research Institute for Synchrotron Radiation Science (HiSOR), Hiroshima University, Higashi-Hiroshima, Hiroshima 739-0046, Japan}

\author{Akio Kimura} 
\affiliation{Graduate School of Advanced Science and Engineering, Hiroshima University, Higashi-Hiroshima, Hiroshima 739-8526, Japan}
\affiliation{International Institute for Sustainability with Knotted Chiral Meta Matter (WPI-SKCM$^2$), Hiroshima University, Higashi-Hiroshima, Hiroshima 739-8526, Japan}
\affiliation{Research Institute for Semiconductor Engineering, Hiroshima University, Higashi-Hiroshima, Hiroshima 739-8527, Japan}

\author{Koji Miyamoto}
\affiliation{Research Institute for Synchrotron Radiation Science (HiSOR), Hiroshima University, Higashi-Hiroshima, Hiroshima 739-0046, Japan}

\author{Taichi Okuda}
\affiliation{International Institute for Sustainability with Knotted Chiral Meta Matter (WPI-SKCM$^2$), Hiroshima University, Higashi-Hiroshima, Hiroshima 739-8526, Japan}
\affiliation{Research Institute for Synchrotron Radiation Science (HiSOR), Hiroshima University, Higashi-Hiroshima, Hiroshima 739-0046, Japan}
\affiliation{Research Institute for Semiconductor Engineering, Hiroshima University, Higashi-Hiroshima, Hiroshima 739-8527, Japan}

\author{Mario Novak} 
\affiliation{Department of Physics, Faculty of Science, University of Zagreb, 10000 Zagreb, Croatia}

\author{Uwe Bovensiepen} 
\affiliation{Faculty of Physics and Center for Nanointegration (CENIDE), University of Duisburg-Essen, 47057 Duisburg, Germany}

\author{Kenta Kuroda} 
\email{kuroken224@hiroshima-u.ac.jp}
\affiliation{Graduate School of Advanced Science and Engineering, Hiroshima University, Higashi-Hiroshima, Hiroshima 739-8526, Japan}
\affiliation{International Institute for Sustainability with Knotted Chiral Meta Matter (WPI-SKCM$^2$), Hiroshima University, Higashi-Hiroshima, Hiroshima 739-8526, Japan}
\affiliation{Research Institute for Synchrotron Radiation Science (HiSOR), Hiroshima University, Higashi-Hiroshima, Hiroshima 739-0046, Japan}
\affiliation{Research Institute for Semiconductor Engineering, Hiroshima University, Higashi-Hiroshima, Hiroshima 739-8527, Japan}

\date{\today}

\begin{abstract}
Understanding coherent optical responses of topological surface states (TSSs) requires disentangling excitation pathways from the electronic band structure.
Here, using angle-resolved two-photon photoemission spectroscopy, we identify two distinct excitation pathways in the TSSs of Bi$_2$Te$_3$: an off-resonant transition via virtual states and a resonant transition via unoccupied intermediate states.
A pronounced modulation of the spectral response is observed, revealing a competition between the two coherent pathways.
This competition is tunable via temperature-induced shifts of the chemical potential, which selectively modify the resonant channel.
These results provide microscopic insight into the optical excitation mechanisms of TSSs and highlight the potential for controlling their optical responses, relevant for future spintronic devices.
\end{abstract}
\maketitle

Over the past two decades, three-dimensional topological insulators (TIs) have gained significant attention as promising materials for spintronics and quantum information technologies~\cite{Hasan10rmp,Ando13jpsj}.
The most distinctive feature of TIs lies in their topological surface states (TSSs), which exhibit Dirac-cone-like energy dispersion~\cite{Zhang2009,Xia2009, Chen2009} and helical spin textures in momentum space~\cite{Okuda2013,Dil2019}.
In particular, a wide range of intriguing optical responses arising from the coupling of light to TSSs has been extensively studied, including colossal Kerr rotation~\cite{THz_kerr_rotation}, photocurrent generation~\cite{McIver2011, Kastl2015, Ogawa2016}, and nonlinear optical responses~\cite{Hsieh_prl2011,Bai2020,schmid_tunable_2021}.
These non-equilibrium optical responses of TSSs provide a promising platform for the optical control of spin-polarized electrons.

To elucidate the mechanism of the ultrafast optical responses of TSSs, it is essential to understand optical coupling involving unoccupied intermediate states above the Fermi energy ($E_{\rm{F}}$).
In this context, angle-resolved two-photon photoemission spectroscopy (2PPE-ARPES) enables energy- and momentum-resolved access to optically excited electrons~\cite{2PPE_book}.
Because multiphoton excitation is intrinsically coherent, 2PPE-ARPES provides direct sensitivity to transient optical coupling between electronic states within an ultrashort optical pulse and the associated dephasing dynamics~\cite{Petek1997, Petek_2PPE_interf}.
In particular, under resonant excitation conditions, multiple excitation pathways can coexist, interact and compete, which can manifest as characteristic spectral modulations~\cite{Ueba2001,Gudde2007,Fano_prl_2PPE,Cui2014} and transient renormalization of the electronic band dispersion~\cite{Reutzel2020}, as previously studied in well-defined model systems, but remaining largely unexplored in TSSs.

In contrast, time-resolved ARPES based on a pump-probe scheme has been widely employed to investigate nonequilibrium population dynamics of TSSs, including carrier relaxation processes~\cite{Sobota_population_prl2012,Hajlaoui2012,Johaness_prb2014,zhu_ultrafast_2015}, spin polarization dynamics~\cite{Cacho_prl_2015,Jozwiak2016,Barriga_spin_2016,Mori2023,kawaguchi_time_2023}, and electron-phonon interactions~\cite{Sobota_prl_2014}.
In these experiments, the photoemission process is governed by one-photon excitation, and the pump pulse prepares the system in a nonequilibrium state.
While pump-probe ARPES primarily probes population dynamics, access to the coherent aspects of optical excitation processes in TSSs is more naturally achieved using multiphoton schemes such as 2PPE-ARPES~\cite{Aeschlimann2025}.
As a result, experimental studies of coherent optical phenomena in TSSs have remained elusive, with only a few observations so far, including resonant photocurrent generations~\cite{kuroda_generation_2016,kuroda_ultrafast_2017,soifer_band-resolved_2019}, lightwave-driven currents~\cite{Reimann2018}, and Floquet--Bloch band formation~\cite{Wang2013,Mahmood2016,Ito2023}.
\begin{figure}[t!]
\begin{center}
\includegraphics[width=1.0\columnwidth]{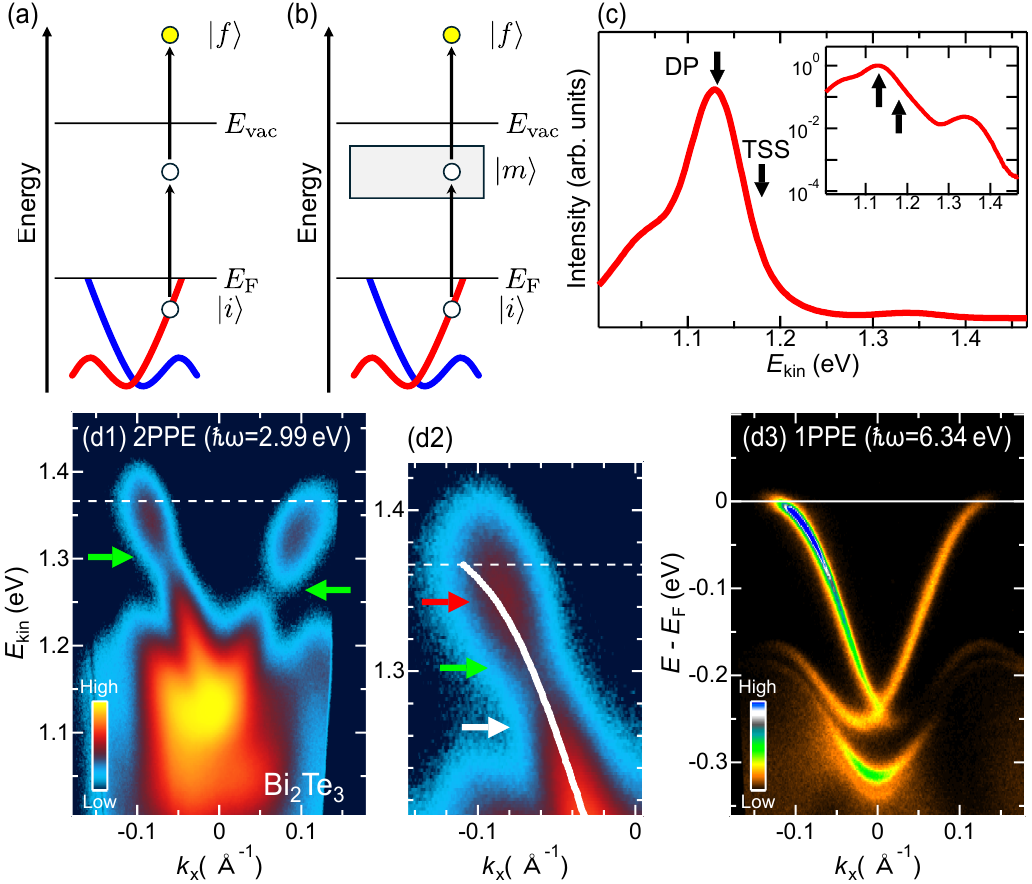}
\caption{
(a, b) Schematic illustration of the two competing paths of single-color 2PPE processes: (a) off-resonant transition from the occupied TSS ($|i>$) to the photoelectron final state ($|f>$) via a virtual state, and (b) resonant transition through unoccupied intermediate states ($|m>$).
(c) Angle-integrated energy distribution curve of the 2PPE spectra from $n$-type Bi$_2$Te$_3$.
The inset shows the same spectrum on a logarithmic scale to highlight the weak signals from the TSS.
The black arrows indicate the TSS-related features, including the Dirac point (DP).
(d1,d2) 2PPE-ARPES results at $T=20$ K, with (d2) a magnified view of (d1); the intensity is displayed on a logarithmic scale.
The dashed line indicates the signal threshold determined from the Fermi-level ($E_{\rm{F}}$) cut-off.
The green arrow marks an intensity node where the 2PPE signal is suppressed.
The white and red arrows in (d2) indicate features assigned to the TSS and the intermediate states, respectively.
White markers in (d2) show the equilibrium TSS dispersion extracted from the 1PPE-ARPES data in (d3) acquired with $\hbar\omega=6.34$ eV~\cite{iwata_spinARPES_2024}.
}
\label{fig1}
\end{center}
\end{figure}

In this Letter, using 2PPE-ARPES, we disentangle two distinct optical excitation pathways of TSSs in Bi$_2$Te$_3$: a resonant excitation via real unoccupied intermediate states and an off-resonant excitation mediated by virtual states.
We elucidate how these two excitation pathways interact and compete in shaping the spectral response, with its pronounced modulations directly reflecting their coherent coupling.
The relative balance between the competing pathways is tunable via temperature-induced shifts of the chemical potential, which selectively modify the resonant channel.
These results establish TSSs as a new platform for coherent and tunable control of optical excitation pathways.

Single-crystalline $n$-type Bi$_2$Te$_3$ and Bi$_2$Se$_3$ were grown by Bridgman method, with Bi$_2$Se$_3$ used as a reference material, and cleaved in situ at room temperature under ultrahigh vacuum conditions better than $1\times10^{-7}$~Pa.
Ultraviolet laser pulses with photon energies ($\hbar\omega$) of 2.99-3.12~eV, a pulse duration of 60~fs, and a pulse energy of approximately 0.2~nJ were used for the 2PPE measurements.
The pulses were generated via second-harmonic generation from a Ti:sapphire oscillator operating at a repetition rate of 76~MHz.
Photoelectrons were detected using a hemispherical electron analyzer (SCIENTA-OMICRON DA30L).
The overall energy and angular resolutions for 2PPE-ARPES were 30~meV and $0.3^{\circ}$, respectively.
The $p$-polarized ultraviolet beam was incident at an angle of $50^{\circ}$ with respect to the surface normal.
The 2PPE-ARPES signals are collected along $\rm{\bar{\Gamma}}$-$\rm{\bar{M}}$ high-symmetry momentum line.
For comparison, conventional one-photon photoemission (1PPE) ARPES measurements were also performed using photons with an energy of $\hbar\omega$=6.34~eV~\cite{iwata_spinARPES_2024}.
All measurements were carried out under ultrahigh vacuum conditions better than $1\times10^{-8}$~Pa.

In 2PPE-ARPES, for $\hbar\omega\sim$3.0~eV used here, a single-photon process is insufficient to emit an electron from an occupied state below $E_{\rm F}$~\cite{2PPE_book}.
Instead, photoemission occurs when the electron absorbs two photons, enabling it to overcome the vacuum level ($E_{\rm{vac}}$).
Two distinct excitation pathways can typically contribute to the 2PPE process.
The first is a direct transition from an occupied initial state to the final state via a virtual intermediate state [Fig.~\ref{fig1}(a)]. 
The 2PPE-ARPES intensity still reflects the occupied band structure, similar to conventional single-photon photoemission.
The second involves a resonant transition through unoccupied intermediate states, which becomes accessible if an appropriate unoccupied band exists [Fig.~\ref{fig1}(b)].
This process allows the 2PPE-ARPES intensity to probe the unoccupied intermediate band structure~\cite{sobota_direct_2013}, offering complementary information to the first pathway.
In practice, both excitation pathways can interact and compete, and the observed 2PPE intensity generally reflects a coherent superposition of these contributions~\cite{Gudde2007,Fano_prl_2PPE,Cui2014}.

\begin{figure*}[t!]
\begin{center}
\includegraphics[width=1.0\textwidth]{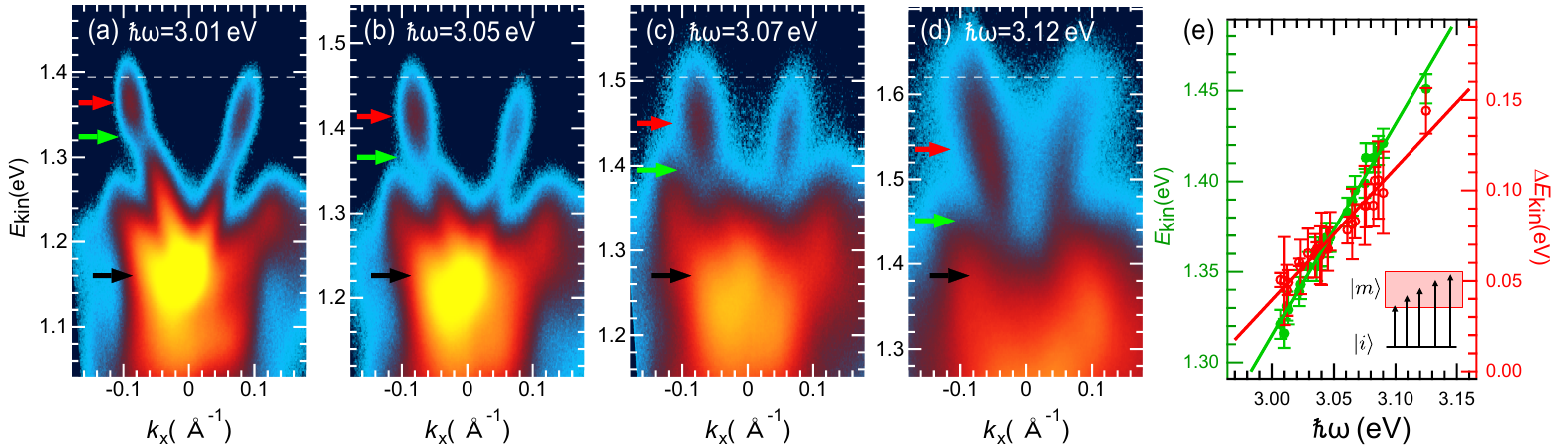}
\caption{
(a)-(d) Representative 2PPE-ARPES results acquired at $T = 20$ K with various excitation $\hbar\omega$.
The dashed line indicates the 2PPE signal threshold determined by the Fermi-level cut-off.
The energies at which the intermediate-state bands, the intensity node, and DP appear are indicated by red, green, and black arrows, respectively.
(e) Plot of the $E_{\rm kin}$ position of the intensity node (green) as a function of $\hbar\omega$, together with the observable energy window of the intermediate-state feature in the 2PPE spectra (red).
The solid lines are linear fits to the experimental data, yielding slopes of 1.1$\pm$0.1 (green) and 0.7$\pm$0.1 (red).
}
\label{fig2}
\end{center}
\end{figure*}

Figure~\ref{fig1}(d1) presents the 2PPE-ARPES result of Bi$_2$Te$_3$ obtained using photons with $\hbar\omega$=2.99~eV at $T$=20~K.
A well-defined Dirac-like dispersion of TSS is clearly observed, with the Dirac point (DP) located at a kinetic energy ($E_{\mathrm{kin}}$) of 1.15~eV.
Since the 2PPE intensity above the DP energy is relatively weak [Fig.~\ref{fig1}(c)], all 2PPE-ARPES results presented hereafter are shown on a logarithmic intensity scale.
We find that signals originating from higher-order multiphoton processes are negligibly small, and therefore a cutoff of the observed 2PPE signals at $E_{\rm{kin}}^{\rm{max}}\sim$1.37~eV [dashed line in Fig.~\ref{fig1}(d1)] originates from $E_{\rm{F}}$ cutoff of the initial state population [see Fig.~\ref{fig1}(a) and \ref{fig1}(b)], which is determined by $E_{\rm{kin}}^{\rm{max}}$=2$\hbar\omega{-}E_{\rm{vac}}-E_{\rm{F}}$.

In particular, an intensity node, where the 2PPE signals are strongly suppressed, is clearly observed in the TSS dispersion [green arrow in Fig.~\ref{fig1}(d1)].
On the negative electron momentum ($k_x$), this node is observed at $E_{\mathrm{kin}}=$1.30~eV, while on the positive $k_x$, it is seen at $E_{\mathrm{kin}}=$1.25~eV.
In Fig.~\ref{fig1}(d2), we further investigate the details of the observed 2PPE dispersion by comparing it with the equilibrium TSS dispersion determined by 1PPE-ARPES with $\hbar\omega$=6.34~eV [Fig.~\ref{fig1}(d3)].
Clearly, the 2PPE dispersion on the lower-energy side of the intensity node shows good agreement with the dispersion obtained from 1PPE-ARPES [white arrow in Fig.~\ref{fig1}(d2)], confirming the TSS dispersion in the 2PPE process [Fig.~\ref{fig1}(a)].
In contrast, the 2PPE dispersion on the higher-energy side substantially deviates from the 1PPE result [red arrow in Fig.~\ref{fig1}(d2)], suggesting the observation of the intermediate states inherent to the 2PPE process.
In contrast to Bi$_2$Te$_3$, our results obtained for Bi$_2$Se$_3$ demonstrate good agreement between the 2PPE and 1PPE dispersions [Supplementary Note I].
Therefore, the deviation observed in the 2PPE signals relative to the 1PPE results represents a characteristic feature specific to Bi$_2$Te$_3$.

To examine whether the 2PPE intensity node in Bi$_2$Te$_3$ is linked to excitation via intermediate states above $E_{\rm{F}}$ [Fig.~1(b)], we study its dependence on $\hbar\omega$ in Figs.~2(a-d). As $\hbar\omega$ is increased from 3.01~eV to 3.13~eV, the energy position of the intensity node (green arrows) gradually shifts toward DP (black arrows).
Since the 2PPE signal near the DP originates from the initial TSS below $E_{\rm{F}}$, the observed $\hbar\omega$ dependence of the intensity node cannot be explained by the initial-state contribution alone [Fig.~1(a)].

To quantify this trend, we plot $E_{\rm{kin}}$ of the intensity node as a function of $\hbar\omega$ [green markers in Fig.~2(e)].
Over $3.00 \le \hbar\omega \le 3.12$~eV, the node energy position shows an approximately linear dependence with a fitted slope of 1.1$\pm$0.1, indicating that it scales nearly linearly with $\hbar\omega$.
Such a near-$\hbar\omega$ scaling is consistent with excitation mediated by real intermediate states [Fig.~1(b)], whereas a purely virtual-state process would lead to a $2\hbar\omega$ scaling of the final-state energy [see Fig.~1(a)].

In addition, as $\hbar\omega$ increases, the energy window over which the intermediate-state feature becomes visible expands [red arrows in Figs.~2(a-d)].
This behavior is naturally explained by the increased accessible energy range of intermediate states under higher-$\hbar\omega$ excitation, which broadens the observable energy window in 2PPE-ARPES [inset of Fig.~2(e)].
To quantify this effect, we plot the spectral width of the intermediate-state feature, $\Delta E_{\rm{kin}}$, as a function of $\hbar\omega$ [red markers in Fig.~2(e)].
A linear fit yields a slope of 0.7$\pm$0.1, again close to an $\hbar\omega$-linear dependence expected for excitation via real intermediate states above $E_{\rm{F}}$ [Fig.~1(b)].
\begin{figure*}[t!]
\begin{center}
\includegraphics[width=1.0\textwidth]{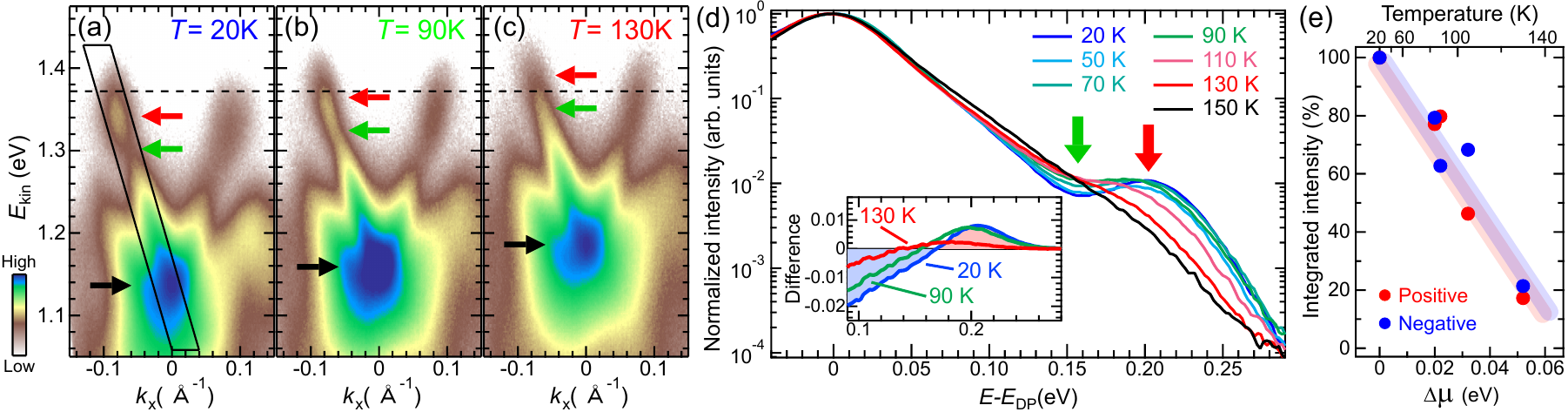}
\caption{
(a)-(c) The representative 2PPE-ARPES results with $\hbar\omega$=3.00~eV at different temperatures.
The energies at which the intermediate-state bands, the intensity node, and DP appear are indicated by red, green, and black arrows, respectively.
(d) The logarithmic plots of the integrated 2PPE signals along the TSS dispersion within the $E$-$k$ window denoted by the parallelogram in (a).
To compensate for the temperature-induced shift of the chemical potential, the horizontal axis is referenced to the DP energy ($E$-$E_{\rm{DP}}$).
The red and green arrows indicate the energy positions of the peak and dip intensities.
The inset shows the differential intensities with respect to the data at $T$=150~K where the 2PPE signals of the intermediate state are negligibly small.
(e) Normalized area intensities of the positive and negative components extracted from the difference spectra [inset of (d)].
The areas are obtained by integrating both parts and are normalized to their respective values at 20~K.
The horizontal axis represents the relative shift of the chemical potential with respect to its position at 20~K, $\Delta{\mu}$, estimated from the $T$-induced shift of $E_{\rm{DP}}$.
The shaded lines are guides to the eye.
}
\label{fig3}
\end{center}
\end{figure*}

On the basis of these experimental findings, we conclude that the 2PPE-ARPES spectra contain two distinct contributions associated with virtual-state and intermediate-state excitation pathways [Fig.~\ref{fig1}(a) and Fig.~\ref{fig1}(b)], which dominate on the lower- and higher-energy sides of the intensity node, respectively.
The intensity modulation observed around the node energy is explained by the coherent simultaneous contribution of these pathways, a hallmark of the ultrafast multiphoton excitation process that is not accessible in conventional one-photon photoemission or pump-probe ARPES.

This conclusion is supported by the temperature dependence of the 2PPE signals.
Figures~\ref{fig3}(a-c) show 2PPE-ARPES results acquired at different temperatures.
Although the TSS dispersion is observed immediately above DP (black arrows), the energy position of DP changes with temperature.
Specifically, at $T$=20~K [Fig.~\ref{fig3}(a)], DP is located at $E_{\mathrm{kin}}\sim$1.15~eV, whereas it gradually shifts toward higher $E_{\mathrm{kin}}$ with increasing temperature, reaching $E_{\rm{kin}}\sim$1.18~eV at $T$=130~K [Fig.~\ref{fig3}(c)].
Since the observed $T$-dependent shift of the DP is reproducible over thermal cycling, it can be attributed to a thermally induced shift of the bulk chemical potential, arising from the asymmetric energy dependence of the density of states at $E_{\mathrm{F}}$\cite{Brouet_prl2013, Song_prb2022, Takeuchi_2011}.

Importantly, together with the shift of the chemical potential, 2PPE-ARPES shows changes in the excitation into the intermediate states.
Although the intermediate-state bands are clearly observed at $T$=20~K [Fig.~\ref{fig3}(a)], the 2PPE signals of the resonant channel gradually weaken with increasing $T$ and eventually disappear at $T$=130~K [red arrow in Fig.~3(c)].
This behavior can be attributed to the shift of the chemical potential toward lower energies at elevated $T$, which leads to depopulation of the initial TSS involved in excitation to the intermediate states, thereby suppressing the corresponding resonant transitions.

To further clarify the suppression of the resonant transition channel, we performed a detailed comparison of the energy distribution curves (EDCs) of the 2PPE intensity along the TSS dispersion at each $T$ [parallelogram region in Fig.~3(a)], which is summarized in Fig.~3(d).
To compensate for the temperature-induced shift of the chemical potential, the horizontal axis in Fig.~3(d) is referenced to the DP energy ($E-E_{\rm DP}$).
At $T$=20~K, a pronounced peak originating from the intermediate states is observed at approximately $E-E_{\rm{DP}}\simeq$0.2~eV.
This peak gradually decreases with increasing temperature and becomes almost invisible at $T$=150~K, where only an exponential intensity distribution associated with the TSS remains.

Notably, the intensity node observed at low temperatures becomes absent at elevated temperatures, where the contribution from the intermediate states is strongly suppressed.
This trend is more clearly visualized in the $T$-dependent difference spectra [inset of Fig.~3(d)], which are obtained by subtracting the EDC collected at $T$=150~K.
At low $T$, an enhancement of the intensity associated with the intermediate states (positive component) develops, while simultaneously an intensity suppression (negative component) emerges on the low-energy side of the node region.
This coupled enhancement-suppression behavior indicates the two excitation pathways influence each other.
The spectral loss at $E-E_{\rm{DP}}$=0.15~eV compared to the high $T$ data rules an incoherent superposition out.
It rather indicates a coherent coupling, arising from interactions and competition between the two excitation channels.

With all these results, we conclude that excitation pathways mediated by virtual states and real intermediate states interact coherently, and that their competition manifests as modulations in the 2PPE spectra intensity.

A question remains as to why the energy position of the intensity node does not coincide for $\pm{k_x}$ in Fig.~\ref{fig1}(d1).
This asymmetry is found to be reversed upon azimuthal rotation of the sample by 180$^\circ$ (Supplementary Note II), demonstrating that it reflects an intrinsic effect associated with the $C_{\rm{3v}}$ symmetry of the Bi$_2$Te$_3$(111) surface.
While the TSS dispersion along the $\bar{\Gamma}$-$\rm{\bar{M}}$ direction itself remains symmetric~\cite{Fu_prl2009}, the $C_{\rm{3v}}$ symmetry of the bulk-derived electronic states~\cite{Chen2009} allows the resonance conditions to differ for $\pm{k_x}$.
As a result, momentum-dependent optical excitation processes arise, giving rise to the observed asymmetry in the spectral modulation.
These observations indicate that the resonant excitation observed here likely originates from transitions from the occupied TSS into bulk-derived unoccupied states.

Figure~\ref{fig3}(e) plots the integrated intensities of the resonant and suppressed components of the difference spectra shown in the inset of Fig.~\ref{fig3}(d), normalized to their values at 20~K.
The horizontal axis represents the relative shift of the chemical potential with respect to its value at 20~K ($\Delta\mu$), estimated from the $T$-dependent shift of $E_{\rm{DP}}$.
The results demonstrate that the resonant excitation channel is strongly activated when the chemical potential satisfies the resonance condition at low temperatures, whereas its contribution is rapidly suppressed as the chemical potential shifts away from this condition at higher $T$.

The linear dependence of the integrated intensity on $\Delta\mu$ indicates that the chemical-potential shift primarily determines the suppression of the resonant channel.
This finding provides strong evidence that the $T$-induced shift of the bulk chemical potential plays a role analogous to an applied bias for TSS, enabling selective on/off control of the resonant excitation pathway.
Building on this analogy, our results naturally point to device-oriented implementations, including optospintronic schemes that exploit light-driven control of spin-polarized surface carriers.
Although gate-controlled optical responses of TSSs have been demonstrated previously~\cite{Duan_sr2014,Kastl_nc2015,Sinha_acs2023,DiGaspare_nature2025}, the present work provides a fundamental understanding of the responsible microscopic mechanism.
The pathway-selective control demonstrated here should be achievable via an external bias or electrostatic gating.

In summary, 2PPE-ARPES allows us to disentangle competing two-photon excitation pathways of TSSs and to control their relative contributions via tuning of the chemical potential.
Since two-photon photoelectron emission itself is a non-linear process, the present results are important not only for understanding coherent nonlinear optical responses of TSSs and their associated dephasing dynamics, but also for realizing their pathway-selective coherent control.
Thereby, our results establish TSSs as a platform for coherent and tunable control of optical excitation pathways, with direct relevance to future optospintronic functionalities.

We thank Kaishu Kawaguchi for fruitful discussions.
This work is supported by JSPS KAKENHI (Grants No. JP21H04652, JP22H01943, No. JP23K17671, No. JP23K23211, and No. JP24KK0107), by the Deutsche Forschungsgemeinschaft (DFG, German Research
Foundation) through Project No. 278162697 (SFB 1242), by JST FOREST Program (Grant No JPMJFR236H and JPMJFR2444), by the Murata Science and Education Foundation, by Yamada Science Foundation, and also by the Collaborative Research Projects of Laboratory for Materials and Structures, Institute of Innovative Research, Science Tokyo.
M.~N. acknowledges support from the Croatian Science Foundation under the Project No. IP-2025-02-3955.
The 2PPE-ARPES measurements were performed with the approval of Hiroshima Synchrotron Radiation Institute (HiSOR) (Proposal No. 25AU004).

\bibliography{bib_arXiv}
\input{SI_arXiv.tex}
\end{document}

%% file: SI_arXiv.tex
\clearpage
\onecolumngrid

\setcounter{figure}{0}
\renewcommand{\thefigure}{S\arabic{figure}}
\renewcommand{\theHfigure}{S\arabic{figure}}
\begin{center}

    {\large \bfseries
    Supplementary Information for\\
    Tunable competing optical excitation pathways\\
    in the topological surface states of Bi$_2$Te$_3$}

    \vspace{1.5em}

    {\normalsize
    Shin Yokoyama,$^{1,\,2}$
    Takahito Takeda,$^{1,\,3}$
    Nagi Suzuki,$^{1,\,2}$
    Takuma Iwata,$^{1,\,2}$
    Ping Zhou,$^{4}$
    Yogendra Kumar,$^{5}$
    Akio Kimura,$^{1,\,2,\,6}$
    Koji Miyamoto,$^{5}$
    Taichi Okuda,$^{2,\,5,\,6}$
    Mario Novak,$^{7}$
    Uwe Bovensiepen,$^{4}$
    and Kenta Kuroda,$^{1,\,2,\,5,\,6}$
    }

    \vspace{1em}

    {\small \itshape
    $^1$Graduate School of Advanced Science and Engineering,\\
    Hiroshima University, Higashi-Hiroshima, Hiroshima 739-8526, Japan\\
    $^2$International Institute for Sustainability with Knotted Chiral Meta Matter (WPI-SKCM$^2$),\\
    Hiroshima University, Higashi-Hiroshima, Hiroshima 739-8526, Japan\\
    $^3$Department of Chemical System Engineering, The University of Tokyo, Bunkyo-ku, Tokyo 113-8656, Japan\\
    $^4$Faculty of Physics and Center for Nanointegration (CENIDE),\\
    University of Duisburg-Essen, 47057 Duisburg, Germany\\
    $^5$Research Institute for Synchrotron Radiation Science (HiSOR),\\
    Hiroshima University, Higashi-Hiroshima, Hiroshima 739-0046, Japan\\
    $^6$Research Institute for Semiconductor Engineering,\\
    Hiroshima University, Higashi-Hiroshima, Hiroshima 739-8527, Japan\\
    $^7$Department of Physics, Faculty of Science, University of Zagreb, 10000 Zagreb, Croatia
    }

    \vspace{2em}
\end{center}

\newcommand{\suppsection}[1]{%
  \par\vspace{1.5em}
  \begin{center}
    {\bfseries #1}
  \end{center}
  \vspace{0.7em}
}

\suppsection{Supplementary Note I: 2PPE-ARPES on Bi$_2$Se$_3$ with $\hbar\omega$=2.99~eV}

\begin{figure}[h!]
\begin{center}
\includegraphics[width=0.6\columnwidth]{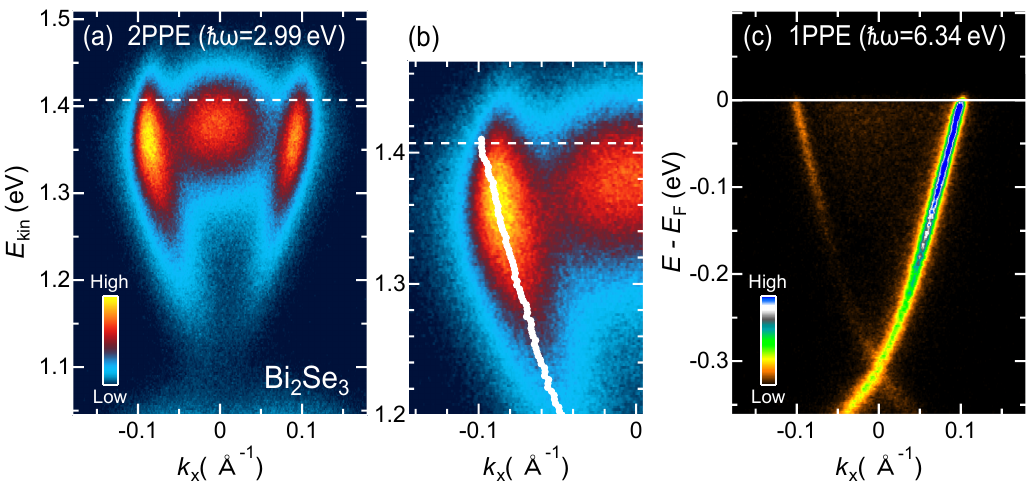}
\caption{
(a) 2PPE-ARPES results of $n$-type Bi$_2$Se$_3$ at $T=20$~K with $\hbar\omega$=2.99~eV.
(b) A magnified view of (a), and the intensity is displayed on a logarithmic scale.
The dashed line indicates the signal threshold determined from the Fermi-level ($E_{\rm{F}}$) cutoff.
The white markers in (b) represent the equilibrium TSS dispersion extracted from the 1PPE-ARPES data in (c), acquired with $\hbar\omega$=6.34~eV.
These data are shown for comparison with the $n$-type Bi$_2$Te$_3$ results presented in Fig.~1 of the main text.
}
\label{figS1}
\end{center}
\end{figure}

As discussed in the main text, we find that the pronounced intensity node observed in 2PPE-ARPES of $n$-type Bi$_2$Te$_3$ originates from a resonant optical excitation between the topological surface state (TSS) and unoccupied intermediate states at excitation photon energies of $\hbar\omega \sim 3$~eV.
This resonance enables coherent coupling between multiple excitation pathways, giving rise to the characteristic node structure in the 2PPE spectral intensity.

For comparison, we present corresponding data for $n$-type Bi$_2$Se$_3$, which is also a prototypical three-dimensional topological insulator hosting a well-defined TSS, similar to Bi$_2$Te$_3$.

In contrast, no resonant interband transition is observed for $n$-type Bi$_2$Se$_3$, even when using $\hbar\omega=2.99$~eV.
Accordingly, the TSS dispersion determined from the 2PPE spectra shows no pronounced deviation from the equilibrium dispersion extracted from 1PPE.
This qualitative difference from Bi$_2$Te$_3$ indicates that the electronic structure of the intermediate states above $E_{\mathrm{F}}$ differs between the two materials, such that coherent coupling of the excitation pathways cannot be established at the same $\hbar\omega$.

\suppsection{Supplementary Note II: Azimuthal dependence of 2PPE-ARPES on Bi$_2$Te$_3$}

\begin{figure}[h!]
\begin{center}
\includegraphics[width=0.6\columnwidth]{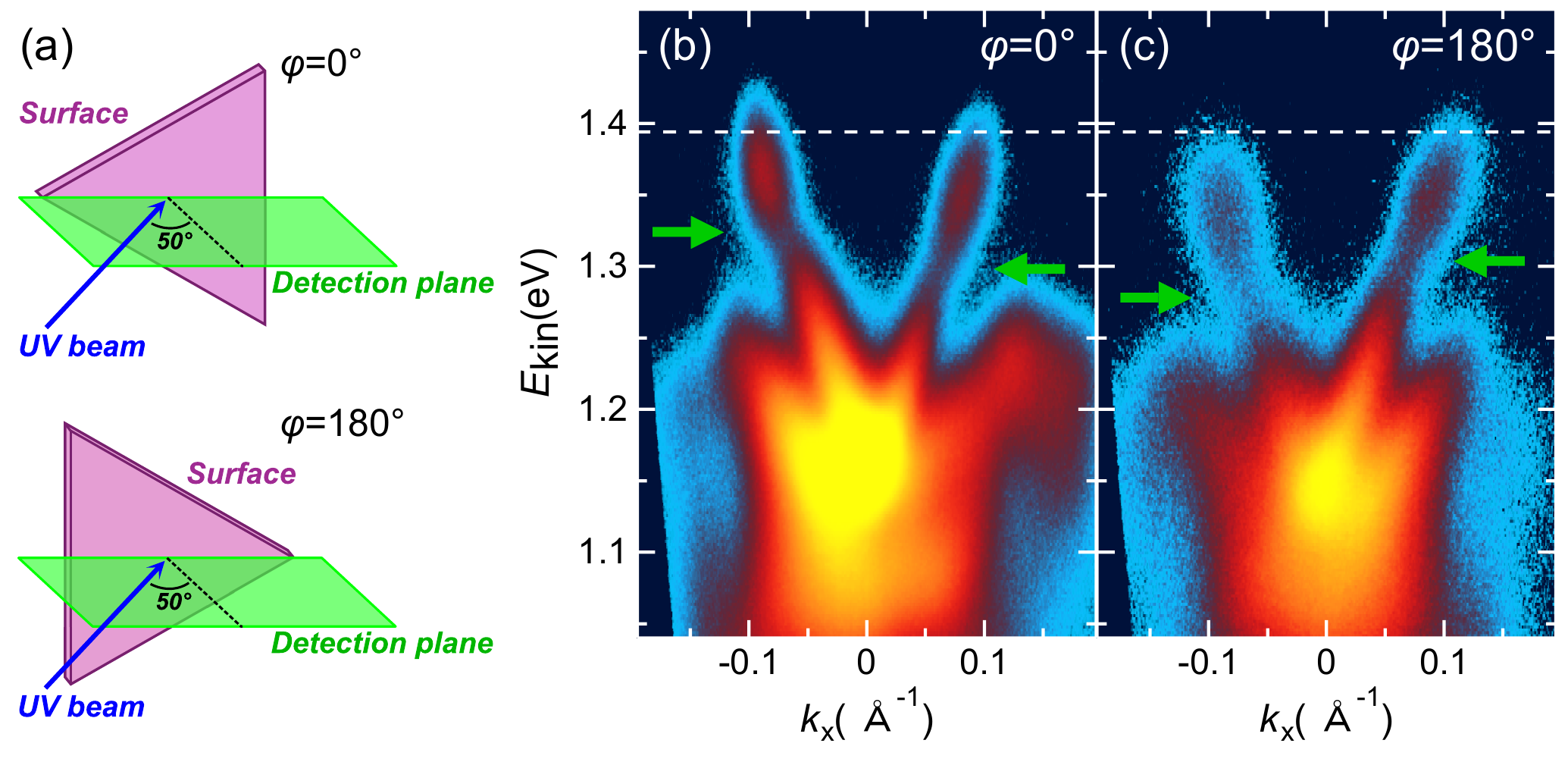}
\caption{
(a) Schematic geometry of the 2PPE-ARPES experiment. Measurements were performed for two sample azimuthal orientations, $\phi$=0$^\circ$ and $\phi$=180$^\circ$. The plane of light incidence is aligned with the photoelectron detection plane, and $p$-polarized light was used. Owing to the $C_{3v}$ surface symmetry of the sample and the measurement along the $\rm{\bar{\Gamma}}$--$\rm{\bar{M}}$ direction, the spectra are asymmetric between $+k_x$ and $-k_x$. This asymmetry is reversed upon rotating the sample by $180^\circ$ in azimuth.
(b,c) 2PPE-ARPES intensity maps of Bi$_2$Te$_3$ acquired at $\phi$=0$^\circ$ and $\phi$=180$^\circ$, respectively, measured at $T=20$~K.
The energy position of the observed intensity node is indicated by the green arrows.
}

\label{figS2}
\end{center}
\end{figure}

The energy position of the intensity node observed in 2PPE-ARPES of $n$-type Bi$_2$Te$_3$ does not coincide between $+k_x$ and $-k_x$ as presented in Fig.~1(d1) in the main text.
This can be attributed to the $C_{3v}$ symmetry of the Bi$_2$Te$_3$(111) surface: along the measured
$\bar{\Gamma}$--$\bar{M}$ direction, $+k_x$ and $-k_x$ are not necessarily symmetry-equivalent.
As a result, the optical excitation matrix elements, as well as the resonance conditions, can differ for the two momenta.

To substantiate this interpretation, we performed an additional experiment in which the same measurement was
repeated after rotating the sample azimuthally by $180^\circ$ [Fig.~S2(a)].
In the dataset obtained in the main text, the node was located at $E_{\mathrm{kin}}=1.32$~eV for $+k_x$
and $E_{\mathrm{kin}}=1.30$~eV for $-k_x$ [green arrows in Fig.~S2(b)].
After the in-plane rotation, the node positions at $\pm k_x$ were reversed, appearing at $E_{\mathrm{kin}}=1.30$~eV
for $+k_x$ and $E_{\mathrm{kin}}=1.32$~eV for $-k_x$ [green arrows in Fig.~S2(c)].
Therefore, the observed asymmetry in the node position reflects the crystal symmetry and, in turn, the electronic
band structure of the unoccupied states located approximately $\hbar\omega$ ($\sim 3$~eV) above $E_{\rm{F}}$ in Bi$_2$Te$_3$.

%% file: bib_arXiv.bib
@book{2PPE_book,
  title = {Dynamics at Solid State Surfaces and Interfaces: Current Developments},
  author = {Bovensiepen, Uwe and Petek, Hrvoje and Wolf, Martin},
  ISBN = {9783527633418},
  url = {http://dx.doi.org/10.1002/9783527633418},
  DOI = {10.1002/9783527633418},
  publisher = {Wiley-VCH},
  address   = {Berlin},
  year = {2010},
}

@article{Hasan10rmp,
  title = {Colloquium: Topological insulators},
  author = {Hasan, M. Z. and Kane, C. L.},
  journal = {Rev. Mod. Phys.},
  volume = {82},
  issue = {4},
  pages = {3045--3067},
  numpages = {0},
  year = {2010},
  month = {Nov},
  publisher = {American Physical Society},
  doi = {10.1103/RevModPhys.82.3045},
  url = {https://link.aps.org/doi/10.1103/RevModPhys.82.3045}
}

@article{Ando13jpsj,
author = {Ando ,Yoichi},
title = {Topological Insulator Materials},
journal = {Journal of the Physical Society of Japan},
volume = {82},
number = {10},
pages = {102001},
year = {2013},
doi = {10.7566/JPSJ.82.102001},
URL = {https://doi.org/10.7566/JPSJ.82.102001},
eprint = {https://doi.org/10.7566/JPSJ.82.102001}
}

@article{Okuda2013,
  title = {Spin- and Angle-Resolved Photoemission of Strongly Spin-Orbit Coupled Systems},
  volume = {82},
  ISSN = {1347-4073},
  url = {http://dx.doi.org/10.7566/JPSJ.82.021002},
  DOI = {10.7566/jpsj.82.021002},
  number = {2},
  journal = {Journal of the Physical Society of Japan},
  publisher = {Physical Society of Japan},
  author = {Okuda,  Taichi and Kimura,  Akio},
  year = {2013},
  month = feb,
  pages = {021002}
}

@article{Zhang2009,
  title = {Topological insulators in Bi$_2$Se$_3$,  Bi$_2$Te$_3$ and Sb$_2$Te$_3$ with a single Dirac cone on the surface},
  volume = {5},
  ISSN = {1745-2481},
  url = {http://dx.doi.org/10.1038/nphys1270},
  DOI = {10.1038/nphys1270},
  number = {6},
  journal = {Nature Physics},
  publisher = {Springer Science and Business Media LLC},
  author = {Zhang,  Haijun and Liu,  Chao-Xing and Qi,  Xiao-Liang and Dai,  Xi and Fang,  Zhong and Zhang,  Shou-Cheng},
  year = {2009},
  month = may,
  pages = {438}
}

@article{Xia2009,
  title = {Observation of a large-gap topological-insulator class with a single Dirac cone on the surface},
  volume = {5},
  ISSN = {1745-2481},
  url = {http://dx.doi.org/10.1038/nphys1274},
  DOI = {10.1038/nphys1274},
  number = {6},
  journal = {Nature Physics},
  publisher = {Springer Science and Business Media LLC},
  author = {Xia,  Y. and Qian,  D. and Hsieh,  D. and Wray,  L. and Pal,  A. and Lin,  H. and Bansil,  A. and Grauer,  D. and Hor,  Y. S. and Cava,  R. J. and Hasan,  M. Z.},
  year = {2009},
  month = may,
  pages = {398}
}

@article{Chen2009,
  title = {Experimental Realization of a Three-Dimensional Topological Insulator ${\mathrm{Bi}}_{2}{\mathrm{Te}}_{3}$},
  volume = {325},
  ISSN = {1095-9203},
  url = {http://dx.doi.org/10.1126/science.1173034},
  DOI = {10.1126/science.1173034},
  number = {5937},
  journal = {Science},
  publisher = {American Association for the Advancement of Science (AAAS)},
  author = {Chen,  Y. L. and Analytis,  J. G. and Chu,  J.-H. and Liu,  Z. K. and Mo,  S.-K. and Qi,  X. L. and Zhang,  H. J. and Lu,  D. H. and Dai,  X. and Fang,  Z. and Zhang,  S. C. and Fisher,  I. R. and Hussain,  Z. and Shen,  Z.-X.},
  year = {2009},
  month = jul,
  pages = {178}
}

@article{schmid_tunable_2021,
	title = {Tunable non-integer high-harmonic generation in a topological insulator},
  volume = {593},
  ISSN = {1476-4687},
  url = {http://dx.doi.org/10.1038/s41586-021-03466-7},
  DOI = {10.1038/s41586-021-03466-7},
  number = {7859},
  journal = {Nature},
  publisher = {Springer Science and Business Media LLC},
  author = {{Schmid,  C. P. and Weigl,  L. and Gr\"{o}ssing,  P. and Junk,  V. and Gorini,  C. and Schlauderer,  S. and Ito,  S. and Meierhofer,  M. and Hofmann,  N. and Afanasiev,  D. and Crewse,  J. and Kokh,  K. A. and Tereshchenko,  O. E. and G\"{u}dde,  J. and Evers,  F. and Wilhelm,  J. and Richter,  K. and H\"{o}fer,  U. and Huber,  R.}},
  year = {2021},
  month = may,
  pages = {385}
}

@article{iwata_spinARPES_2024,
  title = {Laser-based angle-resolved photoemission spectroscopy with micrometer spatial resolution and detection of three-dimensional spin vector},
  volume = {14},
  ISSN = {2045-2322},
  url = {http://dx.doi.org/10.1038/s41598-023-47719-z},
  DOI = {10.1038/s41598-023-47719-z},
  number = {1},
  journal = {Scientific Reports},
  publisher = {Springer Science and Business Media LLC},
  author = {Iwata,  Takuma and Kousa,  T. and Nishioka,  Y. and Ohwada,  K. and Sumida,  K. and Annese,  E. and Kakoki,  M. and Kuroda,  Kenta and Iwasawa,  H. and Arita,  M. and Kumar,  S. and Kimura,  A. and Miyamoto,  K. and Okuda,  T.},
  year = {2024}
  }

@article{Kuroda_generation_2016,
  title = {Generation of Transient Photocurrents in the Topological Surface State of ${\mathrm{Sb}}_{2}{\mathrm{Te}}_{3}$ by Direct Optical Excitation with Midinfrared Pulses},
  author = {Kuroda, K. and Reimann, J. and G\"udde, J. and H\"{o}fer, U.},
  journal = {Phys. Rev. Lett.},
  volume = {116},
  issue = {7},
  pages = {076801},
  numpages = {5},
  year = {2016},
  month = {Feb},
  publisher = {American Physical Society},
  doi = {10.1103/PhysRevLett.116.076801},
  url = {https://link.aps.org/doi/10.1103/PhysRevLett.116.076801}
}

@article{kuroda_ultrafast_2017,
  title = {Ultrafast energy- and momentum-resolved surface Dirac photocurrents in the topological insulator ${\mathrm{Sb}}_{2}{\mathrm{Te}}_{3}$},
  author = {Kuroda, Kenta and Reimann, J. and Kokh, K. A. and Tereshchenko, O. E. and Kimura, A. and G\"udde, J. and H\"{o}fer, U.},
  journal = {Phys. Rev. B},
  volume = {95},
  issue = {8},
  pages = {081103},
  numpages = {5},
  year = {2017},
  month = {Feb},
  publisher = {American Physical Society},
  doi = {10.1103/PhysRevB.95.081103},
  url = {https://link.aps.org/doi/10.1103/PhysRevB.95.081103}
}

@article{kawaguchi_time_2023,
	title = {Time-, spin-, and angle-resolved photoemission spectroscopy with a 1-{MHz} 10.7-{eV} pulse laser},
	volume = {94},
	issn = {0034-6748},
	url = {https://doi.org/10.1063/5.0151859},
	doi = {10.1063/5.0151859},
	number = {8},
	urldate = {2025-05-07},
	journal = {Review of Scientific Instruments},
	author = {Kawaguchi, Kaishu and Kuroda, Kenta and Zhao, Z. and Tani, S. and Harasawa, A. and Fukushima, Y. and Tanaka, H. and Noguchi, R. and Iimori, T. and Yaji, K. and Fujisawa, M. and Shin, S. and Komori, F. and Kobayashi, Y. and Kondo, Takeshi},
	year = {2023},
	pages = {083902},
}

@article{soifer_band-resolved_2019,
	title = {Band-{Resolved} {Imaging} of {Photocurrent} in a {Topological} {Insulator}},
	volume = {122},
	url = {https://link.aps.org/doi/10.1103/PhysRevLett.122.167401},
	doi = {10.1103/PhysRevLett.122.167401},
	number = {16},
	urldate = {2025-05-07},
	journal = {Physical Review Letters},
	author = {Soifer, H. and Gauthier, A. and Kemper, A. F. and Rotundu, C. R. and Yang, S.-L. and Xiong, H. and Lu, D. and Hashimoto, M. and Kirchmann, P. S. and Sobota, J. A. and Shen, Z.-X.},
	month = apr,
	year = {2019},
	pages = {167401},
}

@article{sobota_direct_2013,
	 title = {Direct Optical Coupling to an Unoccupied Dirac Surface State in the Topological Insulator ${\mathrm{Bi}}_{2}{\mathrm{Se}}_{3}$},
  author = {Sobota, J. A. and Yang, S.-L. and Kemper, A. F. and Lee, J. J. and Schmitt, F. T. and Li, W. and Moore, R. G. and Analytis, J. G. and Fisher, I. R. and Kirchmann, P. S. and Devereaux, T. P. and Shen, Z.-X.},
  journal = {Phys. Rev. Lett.},
  volume = {111},
  issue = {13},
  pages = {136802},
  numpages = {5},
  year = {2013},
  month = {Sep},
  publisher = {American Physical Society},
  doi = {10.1103/PhysRevLett.111.136802},
  url = {https://link.aps.org/doi/10.1103/PhysRevLett.111.136802}
}

@article{Sobota_population_prl2012,
  title = {Ultrafast Optical Excitation of a Persistent Surface-State Population in the Topological Insulator ${\mathrm{Bi}}_{2}{\mathrm{Se}}_{3}$},
  author = {Sobota, J. A. and Yang, S. and Analytis, J. G. and Chen, Y. L. and Fisher, I. R. and Kirchmann, P. S. and Shen, Z.-X.},
  journal = {Phys. Rev. Lett.},
  volume = {108},
  issue = {11},
  pages = {117403},
  numpages = {5},
  year = {2012},
  month = {Mar},
  publisher = {American Physical Society},
  doi = {10.1103/PhysRevLett.108.117403},
  url = {https://link.aps.org/doi/10.1103/PhysRevLett.108.117403}
}

@article{Johaness_prb2014,
  title = {Spectroscopy and dynamics of unoccupied electronic states of the topological insulators ${\mathrm{Sb}}_{2}{\mathrm{Te}}_{3}$ and ${\mathrm{Sb}}_{2}{\mathrm{Te}}_{2}\mathrm{S}$},
  author = {Reimann, J. and G\"udde, J. and Kuroda, K. and Chulkov, E. V. and H\"{o}fer, U.},
  journal = {Phys. Rev. B},
  volume = {90},
  issue = {8},
  pages = {081106},
  numpages = {5},
  year = {2014},
  month = {Aug},
  publisher = {American Physical Society},
  doi = {10.1103/PhysRevB.90.081106},
  url = {https://link.aps.org/doi/10.1103/PhysRevB.90.081106}
}

@article{zhu_ultrafast_2015,
	title = {Ultrafast electron dynamics at the {Dirac} node of the topological insulator ${\mathrm{Sb}}_{2}{\mathrm{Te}}_{3}$},
	volume = {5},
	issn = {2045-2322},
	url = {https://doi.org/10.1038/srep13213},
	doi = {10.1038/srep13213},
	number = {1},
	journal = {Scientific Reports},
	author = {Zhu, Siyuan and Ishida, Yukiaki and Kuroda, Kenta and Sumida, Kazuki and Ye, Mao and Wang, Jiajia and Pan, Hong and Taniguchi, Masaki and Qiao, Shan and Shin, Shik and Kimura, Akio},
	month = aug,
	year = {2015},
	pages = {13213},
}

@article{Mori2023,
  title = {Spin-polarized spatially indirect excitons in a topological insulator},
  volume = {614},
  ISSN = {1476-4687},
  url = {http://dx.doi.org/10.1038/s41586-022-05567-3},
  DOI = {10.1038/s41586-022-05567-3},
  number = {7947},
  journal = {Nature},
  publisher = {Springer Science and Business Media LLC},
  author = {Mori,  Ryo and Ciocys,  Samuel and Takasan,  Kazuaki and Ai,  Ping and Currier,  Kayla and Morimoto,  Takahiro and Moore,  Joel E. and Lanzara,  Alessandra},
  year = {2023},
  month = feb,
  pages = {249}
}

@article{Petek_2PPE_interf,
  title = {Optical Dephasing in Cu(111) Measured by Interferometric Two-Photon Time-Resolved Photoemission},
  author = {Ogawa, S. and Nagano, H. and Petek, H. and Heberle, A. P.},
  journal = {Phys. Rev. Lett.},
  volume = {78},
  issue = {7},
  pages = {1339--1342},
  numpages = {0},
  year = {1997},
  month = {Feb},
  publisher = {American Physical Society},
  doi = {10.1103/PhysRevLett.78.1339},
  url = {https://link.aps.org/doi/10.1103/PhysRevLett.78.1339}
}

@article{Gudde2007,
  title = {Time-Resolved Investigation of Coherently Controlled Electric Currents at a Metal Surface},
  volume = {318},
  ISSN = {1095-9203},
  url = {http://dx.doi.org/10.1126/science.1146764},
  DOI = {10.1126/science.1146764},
  number = {5854},
  journal = {Science},
  publisher = {American Association for the Advancement of Science (AAAS)},
  author = {G\"{u}dde,  J. and Rohleder,  M. and Meier,  T. and Koch,  S. W. and H\"{o}fer,  U.},
  year = {2007},
  month = nov,
  pages = {1287}
}

@article{Fano_prl_2PPE,
  title = {Two-State Double-Continuum Fano Resonance at the Si(100) Surface},
  author = {Eickhoff, Christian and Teichmann, Martin and Weinelt, Martin},
  journal = {Phys. Rev. Lett.},
  volume = {107},
  issue = {17},
  pages = {176804},
  numpages = {4},
  year = {2011},
  month = {Oct},
  publisher = {American Physical Society},
  doi = {10.1103/PhysRevLett.107.176804},
  url = {https://link.aps.org/doi/10.1103/PhysRevLett.107.176804}
}

@article{Ito2023,
  title = {Build-up and dephasing of Floquet-Bloch bands on subcycle timescales},
  volume = {616},
  ISSN = {1476-4687},
  url = {http://dx.doi.org/10.1038/s41586-023-05850-x},
  DOI = {10.1038/s41586-023-05850-x},
  number = {7958},
  journal = {Nature},
  publisher = {Springer Science and Business Media LLC},
  author = {Ito,  S. and Sch\"{u}ler,  M. and Meierhofer,  M. and Schlauderer,  S. and Freudenstein,  J. and Reimann,  J. and Afanasiev,  D. and Kokh,  K. A. and Tereshchenko,  O. E. and G\"{u}dde,  J. and Sentef,  M. A. and H\"{o}fer,  U. and Huber,  R.},
  year = {2023},
  month = apr,
  pages = {696}
}

@article{Wang2013,
  title = {Observation of Floquet-Bloch States on the Surface of a Topological Insulator},
  volume = {342},
  ISSN = {1095-9203},
  url = {http://dx.doi.org/10.1126/science.1239834},
  DOI = {10.1126/science.1239834},
  number = {6157},
  journal = {Science},
  publisher = {American Association for the Advancement of Science (AAAS)},
  author = {Wang,  Y. H. and Steinberg,  H. and Jarillo-Herrero,  P. and Gedik,  N.},
  year = {2013},
  pages = {453}
}

@article{Mahmood2016,
  title = {Selective scattering between Floquet-Bloch and Volkov states in a topological insulator},
  volume = {12},
  ISSN = {1745-2481},
  url = {http://dx.doi.org/10.1038/nphys3609},
  DOI = {10.1038/nphys3609},
  number = {4},
  journal = {Nature Physics},
  publisher = {Springer Science and Business Media LLC},
  author = {Mahmood,  Fahad and Chan,  Ching-Kit and Alpichshev,  Zhanybek and Gardner,  Dillon and Lee,  Young and Lee,  Patrick A. and Gedik,  Nuh},
  year = {2016},
  month = jan,
  pages = {306}
}

@article{McIver2011,
  title = {Control over topological insulator photocurrents with light polarization},
  volume = {7},
  ISSN = {1748-3395},
  url = {http://dx.doi.org/10.1038/nnano.2011.214},
  DOI = {10.1038/nnano.2011.214},
  number = {2},
  journal = {Nature Nanotechnology},
  publisher = {Springer Science and Business Media LLC},
  author = {McIver,  J. W. and Hsieh,  D. and Steinberg,  H. and Jarillo-Herrero,  P. and Gedik,  N.},
  year = {2011},
  month = dec,
  pages = {96}
}

@article{Ogawa2016,
  title = {Zero-bias photocurrent in ferromagnetic topological insulator},
  volume = {7},
  ISSN = {2041-1723},
  url = {http://dx.doi.org/10.1038/ncomms12246},
  DOI = {10.1038/ncomms12246},
  number = {1},
  journal = {Nature Communications},
  publisher = {Springer Science and Business Media LLC},
  author = {Ogawa,  N. and Yoshimi,  R. and Yasuda,  K. and Tsukazaki,  A. and Kawasaki,  M. and Tokura,  Y.},
  year = {2016},
  month = jul 
}

@article{Bai2020,
  title = {High-harmonic generation from topological surface states},
  volume = {17},
  ISSN = {1745-2481},
  url = {http://dx.doi.org/10.1038/s41567-020-01052-8},
  DOI = {10.1038/s41567-020-01052-8},
  number = {3},
  journal = {Nature Physics},
  publisher = {Springer Science and Business Media LLC},
  author = {Bai,  Ya and Fei,  Fucong and Wang,  Shuo and Li,  Na and Li,  Xiaolu and Song,  Fengqi and Li,  Ruxin and Xu,  Zhizhan and Liu,  Peng},
  year = {2020},
  month = nov,
  pages = {311}
}

@article{THz_kerr_rotation,
  title = {Terahertz Response and Colossal Kerr Rotation from the Surface States of the Topological Insulator ${\mathrm{Bi}}_{2}{\mathrm{Se}}_{3}$},
  author = {Vald\'es Aguilar, R. and Stier, A. V. and Liu, W. and Bilbro, L. S. and George, D. K. and Bansal, N. and Wu, L. and Cerne, J. and Markelz, A. G. and Oh, S. and Armitage, N. P.},
  journal = {Phys. Rev. Lett.},
  volume = {108},
  issue = {8},
  pages = {087403},
  numpages = {5},
  year = {2012},
  month = {Feb},
  publisher = {American Physical Society},
  doi = {10.1103/PhysRevLett.108.087403},
  url = {https://link.aps.org/doi/10.1103/PhysRevLett.108.087403}
}

@article{Kastl2015,
  title = {Ultrafast helicity control of surface currents in topological insulators with near-unity fidelity},
  volume = {6},
  ISSN = {2041-1723},
  url = {http://dx.doi.org/10.1038/ncomms7617},
  DOI = {10.1038/ncomms7617},
  number = {1},
  journal = {Nature Communications},
  publisher = {Springer Science and Business Media LLC},
  author = {Kastl,  Christoph and Karnetzky,  Christoph and Karl,  Helmut and Holleitner,  Alexander W.},
  year = {2015},
  pages = {6617}
}

@article{Barriga_spin_2016,
  title = {Ultrafast spin-polarization control of Dirac fermions in topological insulators},
  author = {S\'anchez-Barriga, J. and Golias, E. and Varykhalov, A. and Braun, J. and Yashina, L. V. and Schumann, R. and Min\'ar, J. and Ebert, H. and Kornilov, O. and Rader, O.},
  journal = {Phys. Rev. B},
  volume = {93},
  issue = {15},
  pages = {155426},
  numpages = {10},
  year = {2016},
  month = {Apr},
  publisher = {American Physical Society},
  doi = {10.1103/PhysRevB.93.155426},
  url = {https://link.aps.org/doi/10.1103/PhysRevB.93.155426}
}

@article{Jozwiak2016,
  title = {Spin-polarized surface resonances accompanying topological surface state formation},
  volume = {7},
  ISSN = {2041-1723},
  url = {http://dx.doi.org/10.1038/ncomms13143},
  DOI = {10.1038/ncomms13143},
  number = {1},
  journal = {Nature Communications},
  publisher = {Springer Science and Business Media LLC},
  author = {Jozwiak,  Chris and Sobota,  Jonathan A. and Gotlieb,  Kenneth and Kemper,  Alexander F. and Rotundu,  Costel R. and Birgeneau,  Robert J. and Hussain,  Zahid and Lee,  Dung-Hai and Shen,  Zhi-Xun and Lanzara,  Alessandra},
  year = {2016},
  month = oct 
}

@article{Cacho_prl_2015,
  title = {Momentum-Resolved Spin Dynamics of Bulk and Surface Excited States in the Topological Insulator ${\mathrm{Bi}}_{2}{\mathrm{Se}}_{3}$},
  author = {Cacho, C. and Crepaldi, A. and Battiato, M. and Braun, J. and Cilento, F. and Zacchigna, M. and Richter, M. C. and Heckmann, O. and Springate, E. and Liu, Y. and Dhesi, S. S. and Berger, H. and Bugnon, Ph. and Held, K. and Grioni, M. and Ebert, H. and Hricovini, K. and Min\'ar, J. and Parmigiani, F.},
  journal = {Phys. Rev. Lett.},
  volume = {114},
  issue = {9},
  pages = {097401},
  numpages = {6},
  year = {2015},
  month = {Mar},
  publisher = {American Physical Society},
  doi = {10.1103/PhysRevLett.114.097401},
  url = {https://link.aps.org/doi/10.1103/PhysRevLett.114.097401}
}

@article{Hajlaoui2012,
  title = {Ultrafast Surface Carrier Dynamics in the Topological Insulator Bi2Te3},
  volume = {12},
  ISSN = {1530-6992},
  url = {http://dx.doi.org/10.1021/nl301035x},
  DOI = {10.1021/nl301035x},
  number = {7},
  journal = {Nano Letters},
  publisher = {American Chemical Society (ACS)},
  author = {Hajlaoui,  M. and Papalazarou,  E. and Mauchain,  J. and Lantz,  G. and Moisan,  N. and Boschetto,  D. and Jiang,  Z. and Miotkowski,  I. and Chen,  Y. P. and Taleb-Ibrahimi,  A. and Perfetti,  L. and Marsi,  M.},
  year = {2012},
  month = jun,
  pages = {3532}
}

@article{Ueba2001,
  title = {Transition to fixed final states in two-photon photoemission from adsorbate-metal systems},
  volume = {169},
  ISSN = {0169-4332},
  url = {http://dx.doi.org/10.1016/S0169-4332(00)00635-8},
  DOI = {10.1016/s0169-4332(00)00635-8},
  journal = {Applied Surface Science},
  publisher = {Elsevier BV},
  author = {Ueba,  H. and Mii,  T.},
  year = {2001},
  month = jan,
  pages = {63}
}

@article{Cui2014,
  title = {Transient excitons at metal surfaces},
  volume = {10},
  ISSN = {1745-2481},
  url = {http://dx.doi.org/10.1038/nphys2981},
  DOI = {10.1038/nphys2981},
  number = {7},
  journal = {Nature Physics},
  publisher = {Springer Science and Business Media LLC},
  author = {Cui,  Xuefeng and Wang,  Cong and Argondizzo,  Adam and Garrett-Roe,  Sean and Gumhalter,  Branko and Petek,  Hrvoje},
  year = {2014},
  month = jun,
  pages = {505}
}

@article{Reutzel2020,
  title = {Coherent multidimensional photoelectron spectroscopy of ultrafast quasiparticle dressing by light},
  volume = {11},
  ISSN = {2041-1723},
  url = {http://dx.doi.org/10.1038/s41467-020-16064-4},
  DOI = {10.1038/s41467-020-16064-4},
  number = {1},
  journal = {Nature Communications},
  publisher = {Springer Science and Business Media LLC},
  author = {Reutzel,  Marcel and Li,  Andi and Wang,  Zehua and Petek,  Hrvoje},
  year = {2020},
  month = may 
}

@article{Petek1997,
  title = {Femtosecond time-resolved two-photon photoemission studies of electron dynamics in metals},
  volume = {56},
  ISSN = {0079-6816},
  url = {http://dx.doi.org/10.1016/S0079-6816(98)00002-1},
  DOI = {10.1016/s0079-6816(98)00002-1},
  number = {4},
  journal = {Progress in Surface Science},
  publisher = {Elsevier BV},
  author = {Petek,  H. and Ogawa,  S.},
  year = {1997},
  month = dec,
  pages = {239}
}

@article{Sobota_prl_2014,
  title = {Distinguishing Bulk and Surface Electron-Phonon Coupling in the Topological Insulator ${\mathrm{Bi}}_{2}{\mathrm{Se}}_{3}$ Using Time-Resolved Photoemission Spectroscopy},
  author = {Sobota, J. A. and Yang, S.-L. and Leuenberger, D. and Kemper, A. F. and Analytis, J. G. and Fisher, I. R. and Kirchmann, P. S. and Devereaux, T. P. and Shen, Z.-X.},
  journal = {Phys. Rev. Lett.},
  volume = {113},
  issue = {15},
  pages = {157401},
  numpages = {5},
  year = {2014},
  month = {Oct},
  publisher = {American Physical Society},
  doi = {10.1103/PhysRevLett.113.157401},
  url = {https://link.aps.org/doi/10.1103/PhysRevLett.113.157401}
}

@article{Reimann2018,
  title = {Subcycle observation of lightwave-driven Dirac currents in a topological surface band},
  volume = {562},
  ISSN = {1476-4687},
  url = {http://dx.doi.org/10.1038/s41586-018-0544-x},
  DOI = {10.1038/s41586-018-0544-x},
  number = {7727},
  journal = {Nature},
  publisher = {Springer Science and Business Media LLC},
  author = {Reimann,  J. and Schlauderer,  S. and Schmid,  C. P. and Langer,  F. and Baierl,  S. and Kokh,  K. A. and Tereshchenko,  O. E. and Kimura,  A. and Lange,  C. and G\"{u}dde,  J. and H\"{o}fer,  U. and Huber,  R.},
  year = {2018},
  month = sep,
  pages = {396-400}
}

@article{Dil2019,
  title = {Spin- and angle-resolved photoemission on topological materials},
  volume = {1},
  ISSN = {2516-1075},
  url = {http://dx.doi.org/10.1088/2516-1075/ab168b},
  DOI = {10.1088/2516-1075/ab168b},
  number = {2},
  journal = {Electronic Structure},
  publisher = {IOP Publishing},
  author = {Dil,  J Hugo},
  year = {2019},
  month = apr,
  pages = {023001}
}

@article{Aeschlimann2025,
  title = {Time-resolved photoelectron spectroscopy at surfaces},
  volume = {753},
  ISSN = {0039-6028},
  url = {http://dx.doi.org/10.1016/j.susc.2024.122631},
  DOI = {10.1016/j.susc.2024.122631},
  journal = {Surface Science},
  publisher = {Elsevier BV},
  author = {Aeschlimann,  Martin and Bange,  Jan Philipp and Bauer,  Michael and Bovensiepen,  Uwe and Elmers,  Hans-Joachim and Fauster,  Thomas and Gierster,  Lukas and H\"{o}fer,  Ulrich and Huber,  Rupert and Li,  Andi and Li,  Xintong and Mathias,  Stefan and Morgenstern,  Karina and Petek,  Hrvoje and Reutzel,  Marcel and Rossnagel,  Kai and Sch\"{o}nhense,  Gerd and Scholz,  Markus and Stadtm\"{u}ller,  Benjamin and St\"{a}hler,  Julia and Tan,  Shijing and Wang,  Bing and Wang,  Zehua and Weinelt,  Martin},
  year = {2025},
  month = mar,
  pages = {122631}
}

@article{Hsieh_prl2011,
  title = {Nonlinear Optical Probe of Tunable Surface Electrons on a Topological Insulator},
  author = {Hsieh, D. and McIver, J. W. and Torchinsky, D. H. and Gardner, D. R. and Lee, Y. S. and Gedik, N.},
  journal = {Phys. Rev. Lett.},
  volume = {106},
  issue = {5},
  pages = {057401},
  numpages = {4},
  year = {2011},
  month = {Feb},
  publisher = {American Physical Society},
  doi = {10.1103/PhysRevLett.106.057401},
  url = {https://link.aps.org/doi/10.1103/PhysRevLett.106.057401}
}

@article{Takeuchi_2011,
  title={Effect of Chemical Potential on Thermoelectric Power of Bi$_2$Te$_3$ and Bi$_2$Se$_3$},
  author={Akio Yamamoto and Koto Ogawa and Tsunehiro Takeuchi},
  journal={Materials Transactions},
  volume={52},
  number={8},
  pages={1539-1545},
  year={2011},
  doi={10.2320/matertrans.E-M2011809}
}

@article{Brouet_prl2013,
  title = {Large Temperature Dependence of the Number of Carriers in Co-Doped ${\mathrm{BaFe}}_{2}{\mathrm{As}}_{2}$},
  author = {Brouet, V. and Lin, Ping-Hui and Texier, Y. and Bobroff, J. and Taleb-Ibrahimi, A. and Le F\`evre, P. and Bertran, F. and Casula, M. and Werner, P. and Biermann, S. and Rullier-Albenque, F. and Forget, A. and Colson, D.},
  journal = {Phys. Rev. Lett.},
  volume = {110},
  issue = {16},
  pages = {167002},
  numpages = {5},
  year = {2013},
  month = {Apr},
  publisher = {American Physical Society},
  doi = {10.1103/PhysRevLett.110.167002},
  url = {https://link.aps.org/doi/10.1103/PhysRevLett.110.167002}
}

@article{Song_prb2022,
  title = {Temperature dependence of band shifts induced by impurity ionization in ${\text{ZrTe}}_{5}$},
  author = {Song, Wenhua and Zhao, Lingxiao and Wu, Xuchuan and Wang, Zilu and Liu, Qingxin and Li, Man and Ma, Huan and Ding, Pengfei and Gu, G.-D. and Chen, G.-F. and Wang, S.-C.},
  journal = {Phys. Rev. B},
  volume = {106},
  issue = {11},
  pages = {115124},
  numpages = {7},
  year = {2022},
  month = {Sep},
  publisher = {American Physical Society},
  doi = {10.1103/PhysRevB.106.115124},
  url = {https://link.aps.org/doi/10.1103/PhysRevB.106.115124}
}

@article{Fu_prl2009,
  title = {Hexagonal Warping Effects in the Surface States of the Topological Insulator ${\mathrm{Bi}}_{2}{\mathrm{Te}}_{3}$},
  author = {Fu, Liang},
  journal = {Phys. Rev. Lett.},
  volume = {103},
  issue = {26},
  pages = {266801},
  numpages = {4},
  year = {2009},
  month = {Dec},
  publisher = {American Physical Society},
  doi = {10.1103/PhysRevLett.103.266801},
  url = {https://link.aps.org/doi/10.1103/PhysRevLett.103.266801}
}

@article{Duan_sr2014,
author={Duan, Junxi
and Tang, Ning
and He, Xin
and Yan, Yuan
and Zhang, Shan
and Qin, Xudong
and Wang, Xinqiang
and Yang, Xuelin
and Xu, Fujun
and Chen, Yonghai
and Ge, Weikun
and Shen, Bo},
title={Identification of Helicity-Dependent Photocurrents from Topological Surface States in Bi$_2$Se$_3$ Gated by Ionic Liquid},
journal={Scientific Reports},
year={2014},
month={May},
day={08},
volume={4},
number={1},
pages={4889},
issn={2045-2322},
doi={10.1038/srep04889},
url={https://doi.org/10.1038/srep04889}
}

@article{Sinha_acs2023,
author = {Sinha, Aindrila and Mithun, K. P. and Sood, A. K.},
title = {Time-Resolved Second-Harmonic Generation in Topological Insulator Bi$_2$Te$_3$: Competing Contributions from Dirac Surface States, Surface Photovoltage, and Band Bending},
journal = {ACS Photonics},
volume = {10},
number = {11},
pages = {3944-3954},
year = {2023},
doi = {10.1021/acsphotonics.3c00722},
URL = { 
        https://doi.org/10.1021/acsphotonics.3c00722
},
eprint = { 
        https://doi.org/10.1021/acsphotonics.3c00722
}
}

@article{Kastl_nc2015,
author={Kastl, Christoph
and Karnetzky, Christoph
and Karl, Helmut
and Holleitner, Alexander W.},
title={Ultrafast helicity control of surface currents in topological insulators with near-unity fidelity},
journal={Nature Communications},
year={2015},
month={Mar},
day={26},
volume={6},
number={1},
pages={6617},
issn={2041-1723},
doi={10.1038/ncomms7617},
url={https://doi.org/10.1038/ncomms7617}
}

@article{DiGaspare_nature2025,
author={Di Gaspare, Alessandra
and Ghayeb Zamharir, Sara
and Knox, Craig
and Yagmur, Ahmet
and Sasaki, Satoshi
and Salih, Mohammed
and Li, Lianhe
and Linfield, Edmund H.
and Freeman, Joshua
and Vitiello, Miriam S.},
title={Second and third harmonic generation in topological insulator-based van der Waals metamaterials},
journal={Light: Science {\&} Applications},
year={2025},
month={Sep},
day={22},
volume={14},
number={1},
pages={337},
issn={2047-7538},
doi={10.1038/s41377-025-01847-5},
url={https://doi.org/10.1038/s41377-025-01847-5}
}
